\documentclass[12pt, a4paper] {article}

\usepackage{amsfonts}
\usepackage{verbatim}
\usepackage{color}
\usepackage{amsmath}
\usepackage{graphicx}
\usepackage{bm}
\usepackage{amssymb}
\usepackage{cite}
\usepackage{latexsym}
\usepackage{mathrsfs}
\usepackage{textcomp}
\begin{document}
\begin{center}
{\bf\large{On augmented superfield approach to vector Schwinger model} }

\vskip 1.5 cm

{\sf{ \bf Saurabh Gupta$^1$ and R. Kumar$^2$}}\\
\vskip .1cm
{\it $^1$Instituto de F\'\i sica, Universidade de S\~ao Paulo,\\
C. Postal 66318,  05314-970 S\~ao Paulo, SP, Brazil}\\
\vskip .3cm
{\it $^2$Department of Physics \& Astrophysics,\\University of Delhi, New Delhi$-$110 007, India}\\
\vskip .15cm
{E-mails: {\tt guptasaurabh4u@gmail.com; raviphynuc@gmail.com}}

\end{center}

\vskip 1 cm

\noindent
{\bf Abstract:}
We exploit the techniques of Bonora-Tonin superfield formalism to derive the off-shell nilpotent and 
absolutely anticommuting (anti-)BRST as well as (anti-)co-BRST symmetry transformations for the 
$(1+1)$-dimensional (2D) bosonized vector Schwinger model. In the derivation of above symmetries, we
invoke the (dual)-horizontality conditions as well as gauge and (anti-)co-BRST invariant restrictions on 
the superfields that are defined onto the  $(2,2)$-dimensional supermanifold. We provide geometrical 
interpretation of the above nilpotent symmetries (and their corresponding charges). We also express 
the nilpotency and absolute anticommutativity of the (anti-)BRST and (anti-)co-BRST charges within 
the framework of augmented superfield formalism.

\vskip 1.5 cm



\newpage

\section {Introduction}

In order to covariantly  quantize a gauge theory, Becchi-Rouet-Stora-Tyutin (BRST) formalism provides one of the most natural frameworks. 
In BRST formalism, the unitarity and ``quantum" gauge (i.e. BRST) invariance are respected together at any arbitrary order of perturbative 
computations \cite{brs1,brs2,brs3,tyu}. The BRST formalism has find its application in many of the modern theoretical developments in 
the area of quantum field theories and superstring theories. 
Recently, the BRST approach has been applied to construct a Lagrangian for the fermionic higher spin fields \cite{buch}
as well as in the study of massless and massive fields with totally symmetric arbitrary spin \cite{mate} in AdS space. 
Moreover, BRST quantization of the pure spinor superstring has been carried out \cite{joost} and 
its cohomological  aspects have also been established in the pure spinor formalism \cite{nath}.

There are two pivotal properties associated with the (anti-)BRST symmetries (and their corresponding 
charges): (i) nilpotency of order two, and (ii) anticommutativity.
The origin and geometrical interpretation of these abstract mathematical properties are provided 
by the superfield approach to the BRST formalism (see Refs. \citen{lb1,lb2,del1,del2,nak}). 
In this formalism, a given $D$-dimensional gauge theory is generalized onto a $(D, 2)$-dimensional
supermanifold characterized by a pair of the Grassmannian superspace coordinates ($\theta, \bar \theta$). The horizontality condition (HC) 
and gauge invariant restrictions (GIRs)
play a central role in the derivation of the (anti-)BRST symmetry transformations for the gauge and corresponding (anti-)ghost fields 
for a underlying gauge theory. Furthermore, the geometrical basis for the (anti-)BRST symmetries and corresponding generators is 
provided by the translational generators along the Grassmannian directions.

In our present study, we apply extensively the above 
mentioned geometrical superfield formalism to discuss various symmetry properties for the bosonized vector Schwinger model (VSM) within 
the framework of BRST formalism. The VSM, an offshoot of Schwinger model -- quantum electrodynamics in $(1+1)$-dimensions with massless 
fermions, is a well-known 
model in the regime of two-dimensional field theories \cite{sch1,suss,suss1,halp,boyan,raja1,raja2,flack,malik,sgcsm}. It is a gauge 
invariant and exactly solvable model that is endowed with the first-class constraints in the language of Dirac's prescription for the 
classification of constrained systems. This model has been studied within the framework of BRST as well as Hamiltonian formalism
\cite{usha}. In a most recent development, it has been shown that the VSM is equipped with, in totally, {\it six} continuous 
symmetries, namely, (anti-)BRST symmetries, (anti-)co-BRST symmetries, bosonic symmetry and a ghost scale symmetry \cite{gupta}. 
This makes VSM to be a tractable model for the Hodge theory \cite{gupta1,gupta2,rk1,rk2}.

Our present investigation is essential on the following ground. First, to derive the off-shell nilpotent (anti-)BRST and 
(anti-)co-BRST symmetry transformations in a physically intuitive manner by exploiting the power and strength of the 
HC and GIRs (within the framework of superfield formalism). Second, to provide the geometrical origin of the above 
mentioned symmetries (and their corresponding generators) in the language of translational generators along the 
Grassmannian directions of the (2, 2)-dimensional supermanifold, on which, VSM is generalized.

The structure of the paper is organized as follows. In Sec. 2, we briefly recapitulate the constraint structure and gauge symmetry 
for the $(1+1)$-dimensional (2D) bosonized version of VSM. We derive the off-shell nilpotent and absolutely anticommuting (anti-)BRST 
symmetries for VSM, within the framework of augmented superfield formalism, in our Sec. 3. Section 4 is  
devoted to the derivation of nilpotent and absolutely anticommuting (anti-)co-BRST symmetries and their 
geometrical interpretation. In Sec. 5, we capture the nilpotency and absolute anticommutativity properties of the (anti-)BRST as 
well as (anti-)co-BRST charges within the framework of superfield formalism. We also provide the geometrical origin of 
these mathematical properties in this section.  Finally, in Sec. 6, we make some concluding remarks. In Appendix, 
we provide precise proof for the {\it ad hoc} choice of auxiliary field that being made in the Sec. 4.

\section {Preliminaries: constraint structure and gauge symmetries}

We begin with the following Lagrangian density of $2D$ bosonized version of VSM:\footnote{Here, we choose the 
2D flat metric $\eta_{\mu\nu}$ with signature $(+1,-1)$ with the Greek indices 
$\mu, \nu ... = 0, 1$. The 2D Levi-Civita tensor $\epsilon_{\mu\nu} = - \epsilon_{\nu\mu}$ is such that 
$\varepsilon_{01} = +1 = - \varepsilon^{01}$ and it obeys $\epsilon^{\mu\nu} \epsilon_{\mu\nu} = - 2 !, \; 
\epsilon^{\mu\nu} \epsilon_{\mu\lambda} = - \delta^\nu_\lambda$, etc.} \cite{boyan,gupta}
\begin{eqnarray}
{\cal L} &=& - \frac{1}{4} F^{\mu\nu}F_{\mu\nu}  + \frac{1}{2} \partial^\mu \phi \partial_\mu \phi
 - e\,\varepsilon^{\mu\nu} \partial_\mu \phi A_\nu \nonumber\\
&\equiv&  \frac{1}{2} E^2  +  \frac{1}{2} \dot \phi^2 - \frac{1}{2} \phi'^2 + e \dot \phi A_1 - e\phi' A_0,
\end{eqnarray} 
where the overdot and prime on the fields denote the time and space derivatives, respectively. In 2D, the curvature tensor 
$F_{\mu\nu} (= \partial_\mu A_\nu - \partial_\nu A_\mu)$ has only electric field $E$ as its existing component and 
the gauge field $A_\mu$ is topological in nature (in the case of 2D). In the above, $\phi$ is massless bosonic field 
and $e$ represents electric charge as the coupling constant. 

The above Lagrangian density is endowed with two first-class constraints: $\Pi^0 \approx 0$ and $\partial_1\Pi^1  
- e \partial_1 \phi \approx 0$ in Dirac's terminology \cite{gupta}. Here, $\Pi^0$ and  $\Pi^1 (= -F^{01} = E)$ 
are the canonical conjugate momenta with respect to the fields $A_0$ and $A_1$, respectively. The canonical conjugate 
momentum with respect to $\phi$ is $\Pi_{(\phi)} = \dot \phi + eA_1$. We work with the first-order Lagrangian density 
which can be written as follows:
\begin{eqnarray} 
 {\cal L}_f &=& \frac{1}{2} \Big(E^2 -\Pi_\phi^2 - \phi^{\prime 2} - e^2 A_1^2 \Big) + \Pi_\phi \; \dot \phi 
+ e \; \Pi_\phi A_1 - e \; \phi^\prime A_0.  \label{fol}
\end{eqnarray}
It is well known fact that the presence of first-class constraints indicates VSM to be a gauge field theoretic model. 
The most general form of the generator $(G)$ in terms of first-class constraints, which generates the gauge transformations, 
can be given as 
\begin{eqnarray}
 G = \int dx \; \big [\dot \chi \; \Pi^0 - \chi \; (E^\prime - e \phi^\prime) \big], 
\end{eqnarray}
where $\chi (x,t)$ is an infinitesimal local gauge parameter. The local gauge symmetries generated from 
the above generator are given as follows: 
\begin{eqnarray}
\delta A_0 = \dot \chi, \qquad  \delta A_1 =  \chi', \qquad \delta \Pi_{\phi} = e \chi', \qquad
 \delta \phi = 0, \qquad \delta E = 0. \label{gt}
\end{eqnarray}
It can be readily checked that the action integral remains invariant under the above local gauge transformations.

\section{Off-shell nilpotent (anti-)BRST symmetries: superfield approach}

We exploit the standard tools and techniques of Bonora-Tonin's (BT) superfield formalism \cite{lb1,lb2} to derive the off-shell 
nilpotent and absolutely anticommuting (anti-)BRST symmetries for the 2D bosonized version of VSM. 
Before going into the details of superfield formalism, it is worthwhile to mention that the fields $A_0(x,t)$ and $A_1(x,t)$ 
are the functions of spacetime variables $(x, t)$. 
In the physical 2D of spacetime,  we define the exterior derivative ($d$) and one-form connection ($A^{(1)}$) as follows:
\begin{eqnarray}
d = dt \partial_t + dx \partial_x, \qquad A^{(1)} = dt A_0 + dx A_1.
\end{eqnarray}
In the BT superfield formalism, we generalize the exterior derivative and one-form connection to the super exterior 
derivative ($\tilde  d$) and super one-form connection ($ \tilde {\cal A}^{(1)}$) in the superspace which is characterized by, 
in addition to $(x, t)$, a pair of Grassmannian variables $\theta, \bar \theta$ (with $\theta^2 = \bar \theta^2 = 0, 
\; \theta \bar \theta + \bar \theta \theta = 0$) as
\begin{eqnarray}
d \to \tilde  d &=& dt \partial_t + dx \partial_x + d\theta \partial_\theta + d \bar \theta \partial_{\bar \theta}, \nonumber\\
A^{(1)} \to \tilde {\cal A}^{(1)} &=& dt \tilde {\cal A}_0 + dx \tilde {\cal A}_1 + d \theta \bar {\cal F} + d \bar \theta {\cal F},
\end{eqnarray}
where  $\tilde {\cal A}_0, \tilde {\cal A}_1$ are the superfields corresponding to  $A_0, A_1$, respectively 
and $(\bar {\cal F}){\cal F}$ are the superfields corresponding to the (anti-)ghost fields $(\bar C)C$. Now these superfields 
are expanded, in terms of basic and secondary fields of the theory, along the Grassmannian directions in the following fashion:
\begin{eqnarray}
\tilde {\cal A}_0(x,t, \theta, \bar \theta) &=& A_0(x, t) + \theta \bar f_1(x, t) + \bar \theta f_1(x, t) 
+ i\theta \bar \theta B_1(x, t),\nonumber\\
\tilde {\cal A}_1(x,t, \theta, \bar \theta) &=& A_1(x, t) + \theta \bar f_2 (x, t)+ \bar \theta f_2(x, t) 
+ i\theta \bar \theta B_2(x, t),\nonumber\\
 {\cal F} (x,t,  \theta, \bar \theta)&=& C(x, t) + i \theta \bar b_1(x, t) + i \bar \theta b_1(x, t)
 + i\theta \bar \theta s_1(x, t),\nonumber\\
 \bar {\cal F}(x,t, \theta, \bar \theta) &=& \bar C(x, t) + i \theta \bar b_2(x, t) + i \bar \theta b_2(x, t)
 + i\theta \bar \theta s_2(x, t). \label{6}
\end{eqnarray} 
In the above expression, $B_1, B_2, b_1, \bar b_1, b_2, \bar b_2$ are bosonic secondary fields whereas the secondary fields 
$f_1, \bar f_1, f_2, \bar f_2, s_1, s_2 $ are fermionic in nature.

Now, applying the standard technique of horizontality condition (HC) which imposes the following restriction:
\begin{eqnarray}
 d A^{(1)} = \tilde d \tilde {\cal A}^{(1)}. \label{hc}
\end{eqnarray}
In other words, the above condition implies that a `physical' quantity (i.e. $E (= d A^{(1)})$ in the present case) must remain 
unaffected by the presence of Grassmannian variables when the former is generalized onto  $(2, 2)$-dimensional supermanifold. 
Exploiting the above HC (\ref{hc}),  we obtain the following algebraic relationships amongst the basic and secondary fields of 
the theory:
\begin{eqnarray}
&& b_1 = 0, \quad \bar b_2 = 0, \quad s = 0, \quad s_1 = 0, \quad b_2 + \bar b_1 = b,\nonumber\\ 
&& f_1 = \dot C, \quad \bar f_1 = \dot {\bar C}, \quad f_2 = \partial_x C, \quad \bar f_2 = \partial_x \bar C, 
 \nonumber\\
&& B_1 = \dot b_2 = - \dot {\bar b}_1, \quad B_2 = \partial_x b_2 = - \partial_x \bar b_1. \label{8}
\end{eqnarray}
Substituting the above relationships (\ref{8}) into the super-expansion of the superfields [cf. (\ref{6})], we get following explicit 
expansions:
\begin{eqnarray}
\tilde {\cal A}^{(h)}_0(x,t, \theta, \bar \theta) &=& A_0(x, t) + \theta \dot {\bar C}(x, t) + \bar \theta \dot C(x, t)
+ i \theta \bar \theta \dot b(x, t),\nonumber\\
&\equiv& A_0(x, t) + \theta[s_b A_0 (x, t)] + \bar \theta [s_{ab} A_0(x, t)] 
+ \theta \bar \theta[s_b s_{ab} A_0(x, t)],\nonumber\\
\tilde {\cal A}^{(h)}_1(x,t, \theta, \bar \theta) &=& A_1(x, t) + \theta {\bar C'}(x, t) + \bar \theta C'(x, t)
+ i \theta \bar \theta b'(x, t),\nonumber\\
&\equiv& A_1(x, t) + \theta[s_b A_1 (x, t)] + \bar \theta [s_{ab} A_1(x, t)] 
+ \theta \bar \theta[s_b s_{ab} A_1(x, t)],\nonumber\\
{\cal F}^{(h)}(x,t, \theta, \bar \theta) &=& C(x, t) +  \theta[-i b(x, t)],\nonumber\\
&\equiv& C(x, t) + \theta[s_b C (x, t)] + \bar \theta [s_{ab} C(x, t)] 
+ \theta \bar \theta[s_b s_{ab} C(x, t)],\nonumber\\
\bar {\cal F}^{(h)}(x,t, \theta, \bar \theta) &=& \bar C(x, t) + \bar \theta [i b(x, t)]\nonumber\\
&\equiv& \bar C(x, t) + \theta[s_b \bar C (x, t)] + \bar \theta [s_{ab} \bar C(x, t)] 
+ \theta \bar \theta[s_b s_{ab} \bar C(x, t)], \label{9}
 \end{eqnarray} 
where we have chosen $b_2 = b = - \bar b_1$. The superscript $(h)$, in the above expression, denotes the  
superfields obtained after the application of HC.  Thus, the (anti-)BRST symmetry transformations for the gauge field and (anti-)ghost
fields can be easily inferred from the above expansions. Now, in order to derive (anti-)BRST symmetries for the other dynamical 
field $(\phi)$ and corresponding momenta $(\Pi_\phi)$, we have to go beyond the BT superfield formalism. In this connection, it is 
to be noted that the following quantity: 
\begin{eqnarray}
A_1 - \frac{1}{e}\, \Pi_{(\phi)},
\end{eqnarray}
remains invariant under the gauge transformations (4). This GIR serves our purpose in deriving the 
off-shell nilpotent (anti-) BRST transformations of the field $\Pi_{(\phi)}$. For this purpose we generalize 
the GIR in the superspace as follows:
\begin{eqnarray}
\tilde {\cal A}_1^{(h)} - \frac{1}{e}\, \tilde {\bf \Pi}_{(\phi)} = A_1 - \frac{1}{e}\, \Pi_{(\phi)}
\end{eqnarray}
where the superfield $\tilde {\bf \Pi}_{(\phi)}$ is given by
\begin{eqnarray}
\tilde {\bf \Pi}_{(\phi)} (x,t,\theta, \bar \theta)&=& \Pi_{(\phi)}(x, t) + \theta \bar f_3(x, t) + \bar \theta f_3(x, t) 
+ i\theta \bar \theta B_3(x, t), \label{13}
\end{eqnarray}
where the secondary fields  $f_3, \bar f_3$ are fermionic and $B_3$ is bosonic in nature. 
Exploiting (10), (12) and (13), we obtain 
\begin{eqnarray}
f_3 = e \, C', \qquad \bar f_3 = e \,\bar C', \qquad B_3 = e\,b'.
\end{eqnarray}
As a consequence, we can write
\begin{eqnarray}
\tilde {\bf \Pi}_{(\phi)}^{(g,h)}(x,t, \theta, \bar \theta) &=&  \Pi_{(\phi)} (x,t)+ \theta e \bar C'(x,t)
+ \bar \theta e C'(x,t)  + i \theta \bar \theta e b'(x,t)\nonumber\\
&\equiv& \Pi_{(\phi)}(x, t) + \theta[s_b \Pi_{(\phi)} (x, t)] + \bar \theta [s_{ab} \Pi_{(\phi)}(x, t)] \nonumber\\
&+& \theta \bar \theta[s_b s_{ab} \Pi_{(\phi)}(x, t)].
\end{eqnarray} 
The (anti-)BRST symmetry transformations for the gauge and (anti-)ghost fields can be readily deduced from above 
expansions.

It is evident from (4) that the field $\phi(x, t)$ itself gauge invariant. As a consequence, 
the superfield $\tilde {\bf \Phi}(x, t, \theta, \bar \theta)$ corresponding to $\phi (x, t)$ 
has to be independent of the Grassmannian variables $(\theta, \bar \theta).$ This statement can be 
corroborated in the following GIR as 
\begin{eqnarray}
\tilde {\bf \Phi}(x, t, \theta, \bar \theta) = \phi(x,t).
\end{eqnarray}   
Finally, we obtain the following off-shell nilpotent as well as absolutely anticommuting (anti-)BRST transformations for all fields:
\begin{eqnarray}
&&s_b A_0 = \dot C, \;\; s_b A_1 = C', \;\; s_b \Pi_{(\phi)} = eC', \;\; s_b \bar C = ib,\;\; s_b[b, C, \phi] = 0,\nonumber\\
&&s_{ab} A_0 = \dot {\bar C}, \;\, s_{ab} A_1 = \bar C', \;\, s_{ab} \Pi_{(\phi)} = e \bar C', \,\; s_{ab}  C = -ib, \;\,
s_{ab}[b, \bar C, \phi] = 0.  \label{17}
\end{eqnarray}
The above (anti-)BRST transformations are off-shell nilpotent of order two (i.e. $s^2_{(a)b} = 0$) and absolutely anticommuting
(i.e. $s_b\, s_{ab} + s_{ab}\, s_b$ = 0) in nature.

We point out that BRST transformation $(s_b)$ of any generic field $\Omega (x, t)$ is 
equivalent to the translation of the corresponding superfield $\tilde {\bf \Omega} (x, t, \theta, \bar \theta)$ along 
the $\bar \theta$-direction while keeping $\theta$-direction fixed. Similarly, the anti-BRST transformation 
$(s_{ab})$ for any generic field can be obtain by taking the  
translation of superfield along the $\theta$-direction and  keeping $\bar \theta$-direction intact.  
These statements can be mathematically corroborated as
\begin{eqnarray}
s_b \Omega (x,t) = \partial_{\bar \theta} \tilde {\bf \Omega} (x, t, \theta, \bar \theta)\Big|_{\theta = 0}, \qquad 
s_{ab} \Omega (x,t) = \partial_{\theta} \tilde {\bf \Omega} (x, t, \theta, \bar \theta)\Big|_{\bar \theta = 0}, 
\end{eqnarray}
where the generic superfield is given in Eqs. (10), (15) and (16).

Using  basic tenets of the BRST formulation, the gauge-fixed Lagrangian density $({\cal L}_b)$ which respects the above 
off-shell nilpotent and absolutely anticommuting (anti-)BRST transformations can be written as  
\begin{eqnarray}
{\cal L}_b &=& {\cal L}_f - s_b \Big[i \bar C \Big(\dot A_0 - A'_1 + \frac{1}{2}\, b\Big) \Big]\nonumber\\
&\equiv& {\cal L}_f + s_{ab} \Big[i \bar C \Big(\dot A_0 - A'_1 + \frac{1}{2}\, b\Big) \Big]\nonumber\\
&\equiv& {\cal L}_f + b(\dot A_0 - A'_1) + \frac{1}{2}\,b^2 - i \dot {\bar C} \dot C + i\bar C' C', \label{19}
\end{eqnarray}
where ${\cal L}_f$ is the first order Lagrangian density given by (\ref{fol}) above.

\section{(Anti-)co-BRST symmetries: superfield approach}
In order to derive {\it proper} (i.e. off-shell nilpotent and absolutely anticommuting) (anti-)co-BRST symmetries, we shall 
take recourse to the dual horizontality condition (DHC) and the augmented version of superfield approach to BRST 
formalism. The DHC imposes following restriction \cite{rpm1} 
\begin{eqnarray}
 \delta A^{(1)} \; = \; \tilde \delta {\tilde A}^{(1)}, \label{dhc}
\end{eqnarray}
where $\delta (= - \ast d \ast)$ is the co-exterior derivative and $\tilde \delta (= - \ast \tilde d \ast)$ 
represents the super-co-exterior derivative. In other words, the above DHC implies that the gauge-fixing term, 
which is invariant under the (anti-)co-BRST symmetries (see Ref. \cite{gupta} for details), should remain 
unaffected by the presence of Grassmannian variables. Exploiting the above DHC (\ref{dhc}) along with the 
superfield expansions defined in (\ref{6}), we obtain 
\begin{eqnarray}
 \partial_\theta F = 0, \qquad \partial_{\bar\theta} \bar F = 0, \qquad 
 \partial_\theta \bar F + \partial_{\bar\theta} F = 0, 
\end{eqnarray}
which, in turn, yields following relationships amongst the basic and secondary fields of the theory: 
\begin{eqnarray}
 \bar b_1 = 0, \qquad b_2 = 0, \qquad s_1 = s_2 = 0,  \qquad b_1 = {\cal B} = - \bar b_2. 
\end{eqnarray}
Substituting back these relationships into the superexpansion of the superfields (in (\ref{6})), we get
\begin{eqnarray}
 F^{(dh)} (x, t, \theta, \bar \theta)  & = & C (x,t) + \bar\theta \; [i {\cal B} (x,t)] \nonumber\\
 & \equiv & C (x,t) + \theta (s_{ad} C (x,t)) + \bar \theta (s_{d} C (x,t)) + \theta \bar \theta (s_d s_{ad} C(x,t)), \nonumber\\
 \bar F^{(dh)} (x, t, \theta, \bar \theta)  &=&  \bar C (x,t) + \theta \; [- i {\cal B} (x,t)] \nonumber\\
 & \equiv & \bar C (x,t) + \theta (s_{ad} \bar C (x,t)) + \bar \theta (s_{d} \bar C (x,t)) + \theta \bar \theta (s_d s_{ad} \bar C(x,t)). \label{23}
\end{eqnarray}
Here the superscript $(dh)$ represents the superexpansion after the application of DHC. From the above, we can identify 
the (anti-)co-BRST symmetries for the (anti-)ghost fields as listed below:
\begin{eqnarray}
&& s_d \; C = i {\cal B}, \qquad s_d \; \bar C = 0, \qquad s_d \; s_{ad} \; C = 0, \nonumber\\ 
&& s_{ad} \; C = 0, \qquad s_{ad} \; \bar C = - i {\cal B},  \quad s_d \; s_{ad} \; {\bar C} = 0.
\end{eqnarray}
Now, we are free to choose the auxiliary field `${\cal B}$' in terms of basic fields of the theory so that both the 
crucial properties, i.e. the nilpotency and absolute anticommutativity, could be satisfied simultaneously. Thus, we 
choose ${\cal B} = (E - e \phi) \equiv (\dot A_1 - A^\prime_0 - e \phi)$. We provide an explicit proof of this choice 
in Appendix A. However, with this choice of auxiliary field, in terms of basic fields of the theory, we can have only 
on-shell nilpotency. In order to restore the off-shell nilpotency, we linearize the kinetic term of the Lagrangian 
(cf. (\ref{19})) with the help of another auxiliary field, say $\bar b$, i.e. 
$\big (\frac{1}{2} E^2 \equiv {\bar b} E - \frac{1}{2}{\bar b}^2 \big)$ (see Ref. 
\cite{gupta} for details). Thus, using Euler-Lagrange equations of motion for $\bar b$, we get $\bar b = E$.

Therefore, keeping above in mind, Eq. (\ref{23}) yields 
\begin{eqnarray}
 F^{(dh)} (x, t, \theta, \bar \theta)  & = & C (x,t) + \theta \; (0) + \bar\theta \; [i (\bar b - e \phi)] (x,t) + \theta \bar\theta \; (0), \nonumber\\
 \bar F^{(dh)} (x, t, \theta, \bar \theta)  &=&  \bar C (x,t) + \theta \; [ - i(\bar b - e \phi)] (x,t) + \bar \theta \; (0) + \theta \bar\theta \; (0).  
\label{25}
\end{eqnarray}
Hence the (anti-)co-BRST symmetries for the (anti-)ghost fields can be easily identified as
\begin{eqnarray}
&& s_d \; C = i ({\bar b} - e \phi), \qquad s_d \;  {\bar b} = 0, \qquad s_d \; \bar C = 0,  \qquad s_d \; s_{ad} \; C = 0, \nonumber\\ 
&& s_{ad} \; C = 0, \quad s_{ad} \; \bar C = - i (\bar b - e \phi), \quad s_{ad} \; {\bar b} = 0, \quad s_d \; s_{ad} \; {\bar C} = 0, 
\end{eqnarray}
where $s_{(a)d} \; \bar b = 0 $ is obtained due to the nilpotency property $(s_{(a)d}^2 = 0)$.
In order to derive the (anti-)co-BRST symmetries for the basic fields of the theory we have to go beyond the DHC. 
Following from the symmetries of the Lagrangian density (\ref{fol}), we note that following quantities remain invariant under (anti-)co-BRST 
symmetries (see Ref. \cite{gupta} for details):
\begin{eqnarray}
&& s_{(a)d} \; [\dot A_0 - A^\prime_1] = 0, \qquad s_{(a)d} \; [\Pi_{(\phi)} - e A_1] = 0, \nonumber\\
&& s_{(a)d} \; [\phi] = 0, \qquad s_{(a)d} \; [2 e \Pi_{(\phi)} A_1 - \Pi^2_{(\phi)} - e^2 A_1^2] = 0. \label{27}
\end{eqnarray}
These (anti-)co-BRST invariant quantities serve our purpose as they should remain independent of Grassmannian variables $\theta$ and $\bar\theta$ 
when the former are generalized onto the (2, 2)-dimensional supermanifold. Thus, we demand that 
\begin{eqnarray}
&& ({{\cal \dot {\tilde A}}}_0 - \tilde {{\cal A}}^\prime_1) (x,t,\theta,\bar\theta) = (\dot A_0 - A^\prime_1) (x,t), \nonumber\\
&& ({ \tilde {\bf \Pi}_{(\phi)}} - e {\tilde {\cal A}}_1) (x,t,\theta,\bar\theta) = (\Pi_{(\phi)} - e A_1) (x,t), \nonumber\\
&& \tilde {\bf \Phi} (x,t,\theta,\bar\theta) = \phi (x,t). \label{aug}
\end{eqnarray}
The first relation, in the above expression, along with (\ref{6}), immediately implies   
\begin{eqnarray}
 \dot{\bar f}_1 - {\bar f}_2^\prime = 0, \qquad \dot{f}_1 - {f}_2^\prime = 0, \qquad \dot B_1 = B_2^\prime. \label{29}
\end{eqnarray}
Here we make the judicious choice, guided by the basic ingredients of augmented superfield formalism, as $f_1 = - \bar C^\prime, \,
{\bar f}_1 = - C^\prime,\, f_2 = - \dot{\bar C}, \,{\bar f}_2 = - \dot C$. However, we provide a precise proof of this choice in 
Appendix A. With these choices, the superexpansion of superfields (cf. (\ref{6})) can be written as,
\begin{eqnarray}
\tilde {\cal A}^{(as)}_0(x,t, \theta, \bar \theta) &=& A_0(x, t) - \theta {C}^\prime (x, t) - \bar \theta {\bar C}^\prime (x, t) 
- i \theta \bar \theta ({\bar b}^\prime - e \phi^\prime)(x, t),\nonumber\\
\tilde {\cal A}^{(as)}_1(x,t, \theta, \bar \theta) &=& A_1(x, t) - \theta \dot C (x, t) - \bar \theta \dot {\bar C}(x, t) 
- i\theta \bar \theta(\dot {\bar b} - e \dot \phi) (x, t), \label{30}
\end{eqnarray}
where the superscript (as) represents the expansions after the application of the augmented superfield formalism. The second 
relation in (\ref{aug}) along with the inputs from (\ref{13}) implies 
\begin{eqnarray}
 \bar f_3 = e \bar f_2, \qquad f_3 = e f_2, \qquad B_3 = e B_2, 
\end{eqnarray}
which, with the help of (\ref{29}) and (\ref{30}), yields 
\begin{eqnarray}
 \bar f_3 = - e \dot C, \qquad f_3 = - e \dot{\bar C}, \qquad B_3 = - e (\dot {\bar b} - e \dot \phi).
\end{eqnarray}
Thus, Eq. (\ref{13}) reduces to 
\begin{eqnarray}
 \tilde {\bf \Pi}_{(\phi)}^{(as)}(x,t, \theta, \bar \theta) &=&  \Pi_{(\phi)} (x,t) - \theta (e \dot C)(x,t)
- \bar \theta (e \dot {\bar C}) (x,t)  - i \theta \bar \theta e (\dot {\bar b} - e \dot \phi)(x,t)\nonumber\\
&\equiv& \Pi_{(\phi)}(x, t) + \theta[s_d \Pi_{(\phi)} (x, t)] + \bar \theta [s_{ad} \Pi_{(\phi)}(x, t)] \nonumber\\
&+& \theta \bar \theta[s_d s_{ad} \Pi_{(\phi)}(x, t)]. \label{33}
\end{eqnarray}

As, it clear from the (\ref{27}) that the field $\phi(x,t)$ remains invariant under (anti-)co-BRST symmetry transformation. As a
consequence, the (anti-)co-BRST symmetries can be deduced trivially. Thus, we obtain the following  nilpotent as well 
as absolutely anticommuting (anti-)co-BRST transformations for all fields:
\begin{eqnarray}
&& s_d A_0 = - \bar C^\prime, \qquad s_d A_1 = - \dot{\bar C}, \qquad s_d \Pi_{(\phi)} = - e \dot {\bar C}, \nonumber\\
&& s_d C = i (\bar b - e \phi),  \qquad s_d [\bar C,\, b,\, \bar b,\, \phi] = 0, \nonumber\\
&& s_{ad} A_0 = - C^\prime, \qquad s_{ad} A_1 = - \dot{C}, \qquad s_{ad} \Pi_{(\phi)} = - e \dot { C}, \nonumber\\
&& s_{ad} \bar C = - i (\bar b  - e \phi),  \qquad s_{ad} [C, \,b,\, \bar b,\,  \phi] = 0.  \label{34}
\end{eqnarray}
The above (anti-)co-BRST transformations are on-shell nilpotent of order two (i.e. $s_{(a)d}^2 = 0$) and absolutely anticommuting
in nature (i.e. $s_d s_{ad} + s_{ad} s_d = 0$) in nature.

At this juncture, it is worthwhile to mention that co-BRST transformation of any generic field $\Omega (x, t)$ is 
equivalent to the translation of the corresponding superfield $\Omega (x, t, \theta, \bar \theta)$ along the $\bar \theta$-direction 
while keeping $\theta$-direction fixed. Similarly, the anti-co-BRST transformation of any generic field can be obtain by taking the  
translational of the superfield along the $\theta$-direction and  keeping $\bar \theta$-direction intact. Mathematically, 
\begin{eqnarray}
s_d \Omega (x,t) = \partial_{\bar \theta} \Omega (x, t, \theta, \bar \theta)\Big|_{\theta = 0}, \qquad 
s_{ad} \Omega (x,t) = \partial_{\theta} \Omega (x, t, \theta, \bar \theta)\Big|_{\bar \theta = 0},
\end{eqnarray}
where the generic supefields are given in Eqs. (\ref{23}), (\ref{30}) and (\ref{33}).

\section {Nilpotency and absolute anticommutativity: Superfield approach}

In this section, we capture the nilpotency and absolute anticommutativity property of the (anti-)BRST as well as 
(anti-)co-BRST charges within the framework of superfield formalism.  Exploiting the standard techniques of the 
Noether theorem, it is easy to check that the (anti-)BRST symmetries (cf. (\ref{17})) lead to the derivation of following 
nilpotent $(Q_{(a)b}^2 = 0)$ and conserved (anti-)BRST charges $Q_{(a)b}$ as listed below:
\begin{eqnarray}
Q_{ab} = \int dx \, \big[b \,\dot {\bar C} + E \,{\bar C}^\prime + e\, \phi^\prime \,\bar C  \big] \qquad
Q_{b} = \int dx \, \big[b \,\dot C  + E \,C^\prime + e \,\phi^\prime \,C \big]. \label{36}
\end{eqnarray}
The conservation law of above (anti-)BRST charges can be proven with the help of following Euler-Lagrange equations of motion
\begin{eqnarray}
&& \Box C = 0, \qquad  \Pi_\phi = \dot \phi + e A_1, \qquad E^\prime = \dot b + e \phi^\prime, \nonumber\\ 
&& \Box \bar C = 0, \qquad \dot E = b^\prime + e (\Pi_\phi - e A_1), \qquad \dot \Pi_\phi = \phi'' + e A'_0 . 
\end{eqnarray}
At this juncture, it is interesting to note that the (anti-)BRST charges (cf. (\ref{36})) can also be written, up to a surface term, 
in the following form with the help of above mentioned Euler-Lagrange equations of motion:   
\begin{eqnarray}
Q_{ab} =  \int dx \; \big[ b \; \dot{\bar C} - \dot b \; \bar C \big], \qquad Q_{b} = \int dx \; \big[ b \; \dot{C} - \dot b \; C \big].
\end{eqnarray}
It is straightforward to check, with help of (\ref{17}), that the following is true:  
\begin{eqnarray}
&&Q_{ab} = \int dx\, s_{ab}\,\big[ i\,(\bar C\, \dot C - \dot{\bar C}\,C)\big] \,
\equiv \, \int dx \,\big[s_{b} \, (\,i\,\dot{\bar C} \, \bar C)\big], \nonumber\\
&&Q_{b} = \int dx\, s_{b}\,
\big[ -\,i\,(\bar C\, \dot C - \dot{\bar C}\,C)\big]\, \equiv \, \int dx \,\big[s_{ab} \, (- \,i\,\dot C \, C)\big].
\end{eqnarray}
The above (anti-)BRST charges can be expressed in terms of superfields as follows: 
\begin{eqnarray}
Q_{ab} &=& \,i\, \int dx \biggl[\frac{\partial}{\partial\theta}\,\biggl(\bar F^{(h)}(x, \theta, \bar\theta )\, \dot F^{(h)}(x, \theta, \bar\theta)
- \dot {\bar F}^{(h)}(x, \theta, \bar\theta) \, F^{(h)}(x, \theta, \bar\theta)\biggr) \biggr] \bigg|_{\bar\theta = 0}\nonumber\\
&\equiv & \,i\,\int dx \biggl[\int d\theta \, \biggl(\bar F^{(h)}(x, \theta, \bar\theta)\, \dot F^{(h)}(x, \theta, \bar\theta)
- \dot {\bar F}^{(h)}(x, \theta, \bar\theta) \, F^{(h)}(x, \theta, \bar\theta) \biggr)\biggr]\bigg|_{\bar\theta = 0},\nonumber\\
Q_{ab} &=& \,i \int dx \biggl[\frac{\partial}{\partial{\bar\theta}}\,\biggl(\dot{\bar F}^{({h})}(x, \theta, \bar{\theta})\, 
\bar F^{({h})}(x, \theta, \bar{\theta})\biggr)\biggr] \nonumber\\
&\equiv&  i\int dx \biggl[\int  d\bar\theta\,\biggl(\dot {\bar F}^{({h})}(x, \theta, \bar{\theta})\, 
{\bar F}^{({h})}(x, \theta, \bar{\theta})\biggr)\biggr], \nonumber\\
 Q_{b} &=& -\,i\, \int dx \biggl[\frac{\partial}{\partial\bar\theta}\,\biggl( \bar F^{({h})}(x, \theta, \bar{\theta})\, \dot F^{(h)}(x, \theta, \bar{\theta})
- \dot {\bar F}^{(h)}(x, \theta, \bar{\theta})\,  F^{({h})}(x, \theta, \bar{\theta})\biggr)\, \biggr] \bigg|_{\theta = 0} \nonumber\\
&\equiv & -\,i\,\int dx \biggl[\int d\bar\theta \, \biggl(\bar F^{({h})}(x, \theta, \bar{\theta})\, 
\dot F^{(h)}(x, \theta, \bar{\theta})
- \dot {\bar F}^{(h)}(x, \theta, \bar{\theta})\,  F^{({h})}(x, \theta, \bar{\theta})\biggr)\biggr]\bigg|_{\theta = 0},\nonumber\\
Q_{b} &=& -\,i\,\int dx \biggl[\frac{\partial}{\partial\theta}\,\biggl({\dot F}^{(h)}(x, \theta, \bar\theta)\, 
F^{(h)}(x, \theta, \bar\theta)\biggr) \biggr], \nonumber \\
&\equiv &  -\,i\,\int dx \biggl[\,\int \, d\theta\,\biggl({\dot F}^{(h)}(x, \theta, \bar\theta)\, F^{(h)}(x, \theta, \bar\theta)\biggr) \biggr]. \label{40}
\end{eqnarray}
From the above expressions it is clear that $\partial_{\bar\theta} Q_b = 0, \; \partial_\theta Q_{ab} = 0$ 
hold because of the nilpotency properties (i.e. $\partial_{\bar\theta}^2 = 0, \partial_\theta^2 = 0 $) 
of the translational generators $\partial_{\bar\theta}, \partial_\theta$, respectively. This, in turn, 
implies that the nilpotency ($Q_{(a)b}^2 = 0$) of (anti-)BRST charges is encoded in the following observation  
\begin{eqnarray}
&& \partial_{\theta} \, Q_{ab} = 0 \; \Longleftrightarrow \; s_{ab}\, Q_{ab} = 
i\,\{Q_{ab}, \, Q_{ab}\} = 0 \;  \Longrightarrow \quad Q^2_{ab} = 0, \nonumber\\
&& \partial_{\bar\theta} \, Q_b = 0 \; \Longleftrightarrow \; s_b\, Q_b = i\,\{ Q_b, \, Q_{b}\} = 0 
\; \Longrightarrow \quad  Q^2_b = 0. 
\end{eqnarray}
At this juncture, we would like to point out that $\partial_\theta Q_b = 0$ and  $\partial_{\bar\theta} Q_{ab} = 0$ 
is also true, which can be translated in the language of (anti-)BRST symmetry transformations as follows: 
\begin{eqnarray}
&& \partial_{\bar\theta} \, Q_{ab} = 0 \; \Longleftrightarrow \; s_{b}\, Q_{ab} = i\,\{Q_{ab}, \, Q_b\} = 0 \;
\Longrightarrow \quad  Q_{ab}\, Q_b + Q_b \, Q_{ab} = 0, \nonumber\\
&& \partial_{\theta} \, Q_b = 0 \; \Longleftrightarrow \; s_{ab}\, Q_b = i\,\{Q_b, \, Q_{ab}\} = 0 \;
\Longrightarrow \quad Q_b \, Q_{ab} + Q_{ab}\, Q_b = 0. 
\end{eqnarray}
Thus, we have been able to capture the nilpotency and absolute anticommutativity of (anti-)BRST charges in the 
language of augmented superfield formalism.

Similarly, we can also capture the nilpotency and absolute anticommutativity of the (anti-)co-BRST charges. 
For this purpose, we start with the (anti-)co-BRST charges, which can be expressed as
\begin{eqnarray}
&& Q_{ad} = - \int  dx \; \big[ (\bar b - e \phi) \dot{C} +  b  C^\prime \big] 
\equiv - \int  dx \; \big[ (\bar b - e \phi) \dot{C} -  (\dot {\bar b} - e \dot \phi)  C \big], \nonumber\\ 
&& Q_d =  - \int  dx \; \big[ (\bar b - e \phi) \dot{\bar C}  + b \bar C^\prime  \big] 
\equiv - \int  dx \; \big[ (\bar b - e \phi) \dot{\bar C} - (\dot {\bar b} - e \dot \phi) \bar C \big]. 
\end{eqnarray}
These (anti-)co-BRST charges can also be written in the following more convenient way such that the nilpotency and 
absolute anticommutativity becomes clear and easy to check:
\begin{eqnarray}
&& Q_{ad} = i\, \int dx \,s_{ad} \,\bigl[C\, \dot{\bar C} + \bar C \, \dot C \bigr]
\equiv -\,i\, \int dx\, s_d \, \bigl (\dot C\, C \bigr), \nonumber\\
&& Q_d = \,i\, \int dx \,s_d\, \bigl[C\, \dot{\bar C} + \bar C \, \dot C \bigr]
\equiv \,i\, \int dx \,s_{ad} \,\bigl (\dot {\bar C}\, \bar C \bigr).
\end{eqnarray}
In terms of superfield, within the framework of augmented superfield formalism, these (anti-)co-BRST charges can also be written in 
the following manner:
\begin{eqnarray}
Q_{ad} &=& \int dx \biggl[\frac{\partial}{\partial\theta}\,\biggl(i \bar F^{(dh)}(x, \theta, \bar\theta)\, \dot F^{(dh)}(x, \theta, \bar\theta) 
- i\,\dot{\bar F} ^{(dh)}(x, \theta, \bar\theta)\, F^{(dh)}(x, \theta, \bar\theta)\biggr) \, \biggr]\bigg|_{\bar\theta = 0} \nonumber\\
&\equiv & \int dx \biggl[\int d\theta \, \biggl(i\bar{F}^{(dh)}(x, \theta, \bar\theta)\dot F^{(dh)}(x, \theta, \bar\theta) 
- i\,\dot{\bar F}^{(dh)}(x, \theta, \bar\theta)\, F^{(dh)}(x, \theta, \bar\theta)\biggr)\biggr]\bigg|_{\bar\theta = 0},\nonumber\\
Q_{ad} &=& -\,i\,\int dx \biggl[\frac{\partial}{\partial\bar\theta}\, \biggl(\dot F^{(dh)}(x, \theta, \bar{\theta})\, 
F^{(dh)}(x, \theta, \bar{\theta}) \biggr)\, \biggr] \nonumber \\
&\equiv &  -\,i\,\int dx \biggl[\int\,d\bar \theta\, \biggl(\dot F^{(dh)}(x, \theta, \bar{\theta})\, 
F^{(dh)}(x, \theta, \bar{\theta})\biggr)\, \biggr], \nonumber\\
 Q_d &=& \int dx \,\biggl[\frac{\partial}{\partial{\bar\theta}}\,\biggl(i\, \bar F^{(dh)}(x, \theta, \bar{\theta})\,\dot F^{(dh)}(x, \theta, \bar{\theta})
- i\, \dot{\bar F}^{(dh)}(x, \theta, \bar{\theta})\,F^{(dh)}(x, \theta, \bar{\theta})\biggr)\,\biggr]\bigg|_{\theta = 0}\nonumber\\
&\equiv & \int dx \biggl[\int d\bar\theta\,\biggl(i \bar F^{(dh)}(x, \theta, \bar{\theta})\dot F^{(dh)}(x, \theta, \bar{\theta}) 
- i\, \dot{\bar F}^{(dh)}(x, \theta, \bar{\theta})\,F^{(dh)}(x, \theta, \bar{\theta})\biggr)\,\biggr]\bigg|_{\theta = 0},\nonumber\\
Q_{d} &=& \int dx \,\biggl[\frac{\partial}{\partial{\theta}}\,\biggl(i\, \dot{\bar F}^{(dh)}(x, \theta, \bar\theta)\,
{\bar F}^{(dh)}(x, \theta, \bar\theta)\biggr)\,\biggr] \nonumber \\
&\equiv &  \int dx \biggl[\int d\theta\,\biggl(i\, \dot{\bar F}^{(dh)}(x, \theta, \bar\theta)\,{\bar F}^{(dh)}(x, \theta, \bar\theta)\biggr)\,\biggr]. 
\end{eqnarray}
The nilpotency and absolute anticommutativity of (anti-)co-BRST charges are straightforward to check. For instance 
\begin{eqnarray}
&& \partial_{\theta} \, Q_{ad} = 0 \; \Longleftrightarrow \; s_{ad}\, Q_{ad} = i\,\{Q_{ad}, \, Q_{ad}\} \; \Longrightarrow \; Q^2_{ad} = 0, \nonumber\\
&& \partial_{\bar\theta} \, Q_d = 0 \; \Longleftrightarrow \; s_d\, Q_d = i\,\{Q_d, \, Q_{d}\} \; \Longrightarrow \;  Q^2_d = 0,
\end{eqnarray}
is true because of the nilpotency (i.e. $\partial_\theta^2 =  \partial_{\bar\theta}^2 = 0 $) of the translation generators. 
This precisely proves the nilpotency ($Q_{(a)d}^2 = 0$) of the (anti-)co-BRST charges. For the proof of anticommutativity, 
it is interesting to note that following expressions also hold, namely,
\begin{eqnarray}
 &&\partial_{\theta} \, Q_d = 0 \; \Longleftrightarrow \; s_{ad}\, Q_d = i\,\{Q_d, \, Q_{ad}\} = 0 \;
\Longrightarrow \; Q_d \, Q_{ad} + Q_{ad}\, Q_d = 0,  \nonumber\\
&&\partial_{\bar\theta} \, Q_{ad} = 0 \; \Longleftrightarrow \; s_{d}\, Q_{ad} = i\,\{Q_{ad}, \, Q_d\} = 0 \;
\Longrightarrow \; Q_{ad}\, Q_d + Q_d \, Q_{ad} = 0,
\end{eqnarray}
due to nilpotency ($\partial_\theta^2 =  \partial_{\bar\theta}^2 = 0 $) and absolute anticommutativity properties 
($\partial_\theta \partial_{\bar \theta} + \partial_{\bar \theta} \partial_{\theta} = 0 $) of the translational generators. 
This clearly demonstrate the absolute anticommutativity properties of (anti-)co-BRST charges within the framework of augmented
superfield formalism. Thus, we conclude that the nilpotency and absolute anticommutativity properties of (anti-)BRST as well as
(anti-)co-BRST charges are connected with such properties of translational generators along the Grassmannian directions of the 
$(2,2)$-dimensional supermanifold.

\section{Conclusions}
The central results of our present investigation are the precise derivation of the {\it proper} (i.e. nilpotent and 
absolutely anticommuting) (anti-)BRST as well as (anti-)co-BRST symmetry transformations for the 2D bosonized version of vector Schwinger 
model within the framework of augmented superfield formalism. We have made use of the horizontality condition (HC) and 
gauge invariant restriction to derive the (anti-)BRST symmetries for all the fields of the underlying theory. Additionally, 
in order to derive the complete set of (anti-)co-BRST symmetries, for all the fields of the present theory,
we have exploited the power and strength of dual-HC condition and (anti-)co-BRST invariant restrictions.

We have provided the geometrical origin of the above mentioned continuous symmetries (and their corresponding generators) in the 
language of translation generators along the Grassmannian directions of the $(2,2)$-dimensional supermanifold, on which VSM is generalized.  
Furthermore, within the framework of augmented superfield formalism, we expressed the (anti-)BRST and (anti-)co-BRST charges 
in various forms. Subsequently, we have been able to capture the nilpotency and absolute anticommutativity of above mentioned 
charges in the framework of superfield formalism. We have also shown that the nilpotency and absolute anticommutativity 
are connected with such properties of translational generators along the Grassmannian directions of the $(2,2)$-dimensional supermanifold.

\section*{Acknowledgments}
The research work of S. Gupta is supported by the Conselho Nacional de Desenvolvimento 
Cient\'{i}fico e Tecnol\'{o}gico (CNPq), Brazil, Grant No. 151112/2014-2. R. Kumar would like 
to gratefully acknowledge the financial support from UGC, Government of India, New Delhi,  under PDFSS scheme.

\section*{Appendix A. On the specific choice of auxiliary field} 
In this appendix, we provide explicit derivation of the ad hoc choice made in Sec. 4 for the auxiliary field 
\begin{eqnarray}
{\cal  B} \; = \; E - e \phi \;  \equiv \; \dot A_1 - A_0^\prime - e \phi,
\end{eqnarray}
which ensues the relations $B_1 = - i ({\bar b}^\prime - e \phi^\prime), \; B_2 = - i (\dot {\bar b} - e \dot\phi)$ and also connected 
with the choices $f_1 = - \bar C^\prime, \; {\bar f}_1 = - C^\prime, \; f_2 = - \dot{\bar C}, \; {\bar f}_2 = - \dot C$ 
(cf. comment after (\ref{29})).
This choice is guided by the basic ingredients of superfield formalism according to which all the (anti-)BRST (and/or (anti-)co-BRST) 
invariant quantities should remain independent of the Grassmannian variables $\theta$ and $\bar \theta$ when former are generalized on 
to the supermanifold. Following the above logic, we note that the following quantities:
\begin{eqnarray}
 K_1 = A_1 \dot{\bar C}, \qquad K_2  = A_1 \dot C, \qquad K_3  = A_0 \bar C^\prime, \qquad K_4  = A_0 C^\prime,
\end{eqnarray}
remain invariant under (anti-)co-BRST symmetry transformations (i.e. $s_d K_1 = 0, \; s_{ad} K_2 = 0, \; s_d K_3 = 0, \; s_{ad} K_4 = 0$). 
Thus, keeping above in mind, we demand that 
\begin{eqnarray}
 {\tilde {\cal A}}_1 (x, t, \theta, \bar\theta) \; \dot{\bar F}^{(dh)} (x, t, \theta, \bar\theta) \; 
= \; A_1 (x,t) \; \dot{\bar C} (x,t), \nonumber\\
 {\tilde {\cal A}}_1 (x, t, \theta, \bar\theta) \; \dot{F}^{(dh)} (x, t, \theta, \bar\theta) \; 
= \; A_1 (x,t) \; \dot{ C} (x,t), \nonumber\\
 {\tilde {\cal A}}_0 (x, t, \theta, \bar\theta) \; {\bar F}^{\prime (dh)} (x, t, \theta, \bar\theta) \; 
= \; A_0 (x,t) \; {\bar C}^\prime (x,t), \nonumber\\
 {\tilde {\cal A}}_0 (x, t, \theta, \bar\theta) \; {F}^{\prime(dh)} (x, t, \theta, \bar\theta) \; 
= \; A_0 (x,t) \; {C}^\prime (x,t),
\end{eqnarray}
which lead to the following relationships: 
\begin{eqnarray}
&& \bar f_2 \dot{\bar C} - i A_1 \dot {\cal B} = 0, \qquad f_2 \dot{\bar C} = 0, 
\qquad f_2 \dot {\cal B} - B_2 \dot{\bar C} = 0, \nonumber\\ 
&& f_2 \dot{C} +  i A_1 \dot {\cal B} = 0, \qquad \bar f_2 \dot{C} = 0, 
\qquad \bar f_2 \dot {\cal B} - B_2 \dot{ C}= 0, \nonumber\\ 
&& \bar f_1 {\bar C}^\prime - i A_0  {\cal B}^\prime = 0, \qquad f_1 {\bar C}^\prime = 0, 
\qquad f_1 {\cal B}^\prime - B_1 {\bar C}^\prime = 0, \nonumber\\ 
&& f_1 {C}^\prime + i A_0  {\cal B}^\prime = 0, \qquad \bar f_1 {C}^\prime = 0, 
\qquad \bar f_1 {\cal B}^\prime - B_1 {C}^\prime = 0.
\end{eqnarray}
These relationships immediately imply that $f_1 \propto \bar C^\prime, \; \bar f_1 \propto C^\prime, 
\; f_2 \propto \dot{\bar C}, \; \bar f_2 \propto \dot C, \; B_1 \propto  {\cal B}^\prime, \; B_2 \propto  \dot {\cal B}$. 
We have made similar choices (cf. Sec. 4) with a minus sign for algebraic convenience. Thus, we have derived all the 
secondary fields in term of basic fields of the theory which leads to the derivation of full set of (anti-)co-BRST symmetries.

\end{document}